\documentclass[draft]{agujournal2019}
\usepackage{url} 
\usepackage{lineno}
\usepackage[inline]{trackchanges} 
\usepackage{soul}

\draftfalse

\journalname{JGR: Planets}

\usepackage{pdflscape,epstopdf,textgreek,tabularx,colortbl}
\usepackage[shortlabels]{enumitem}
\newcolumntype{L}{>{\raggedright\arraybackslash}X}
\begin{document}


\title{Possible Transient Luminous Events observed in Jupiter's upper atmosphere}


\authors{Rohini S. Giles\affil{1}, Thomas K. Greathouse\affil{1}, Bertrand Bonfond\affil{2}, G. Randall Gladstone\affil{1,3}, Joshua A. Kammer\affil{1}, Vincent Hue\affil{1}, Denis C. Grodent\affil{3}, Jean-Claude G\'{e}rard\affil{3}, Maarten H. Versteeg\affil{1}, Michael H. Wong\affil{4,5}, Scott J. Bolton\affil{1}, John E. P. Connerney\affil{6,7} and Steven M. Levin\affil{8}}

\affiliation{1}{Southwest Research Institute, San Antonio, Texas, USA}
\affiliation{2}{STAR Institute, LPAP, Universit\'{e} de Li\`{e}ge, Li\`{e}ge, Belgium}
\affiliation{3}{University of Texas at San Antonio, San Antonio, Texas, USA}
\affiliation{4}{University of California, Berkeley, California, USA}
\affiliation{5}{SETI Institute, Mountain View, California, USA}
\affiliation{6}{Space Research Corporation, Annapolis, Maryland, USA}
\affiliation{7}{Goddard Space Flight Center, Greenbelt, Maryland, USA}
\affiliation{8}{Jet Propulsion Laboratory, Pasadena, California, USA}

\correspondingauthor{R. S. Giles}{rgiles@swri.edu}

\begin{keypoints}
\item 11 transient bright flashes were observed by the ultraviolet instrument on the Juno mission
\item The flashes have an average duration of 1.4 ms, are located 260 km above the 1-bar level and are dominated by H\textsubscript{2} emission
\item The observations are consistent with Transient Luminous Events which occur in the upper atmosphere in response to tropospheric lightning
\end{keypoints}


\begin{abstract}

11 transient bright flashes were detected in Jupiter's atmosphere using the UVS instrument on the Juno spacecraft. These bright flashes are only observed in a single spin of the spacecraft and their brightness decays exponentially with time, with a duration of $\sim$1.4 ms. The spectra are dominated by H\textsubscript{2} Lyman band emission and based on the level of atmospheric absorption, we estimate a source altitude of 260 km above the 1-bar level. Based on these characteristics, we suggest that these are observations of Transient Luminous Events (TLEs) in Jupiter's upper atmosphere. In particular, we suggest that these are elves, sprites or sprite halos, three types of TLEs that occur in the Earth's upper atmosphere in response to tropospheric lightning strikes. This is supported by visible light imaging, which shows cloud features typical of lightning source regions at the locations of several of the bright flashes. TLEs have previously only been observed on Earth, although theoretical and experimental work has predicted that they should also be present on Jupiter.
    
\end{abstract}

\section*{Plain Language Summary}

The Juno spacecraft has been in orbit around Jupiter since 2016. One of the instruments on this spacecraft is an ultraviolet spectrograph, which is primarily used to make ultraviolet images of Jupiter's auroras. During the first four years of the mission, the ultraviolet spectrograph has observed 11 transient bright flashes. These bright flashes look similar to lightning, but are located much higher in the atmosphere than the cloudy regions of Jupiter where lightning is generated. We suggest that these are observations of Transient Luminous Events (TLEs) in Jupiter's upper atmosphere. In particular, we suggest that these are elves, sprites or sprite halos, three types of TLEs that produce spectacular flashes of light very high in the Earth's atmosphere in response to lightning strikes between clouds or between clouds and the ground. TLEs have previously only been observed on Earth, although theoretical and experimental work has predicted that they should also be present on other planets, including Jupiter. Comparing and contrasting TLE observations between Jupiter and Earth will improve our understanding of electrical activity in planetary atmospheres. 


\section{Introduction}
\label{sec:intro}

Lightning in Jupiter's atmosphere was first observed with instruments on the Voyager 1 spacecraft; the imaging camera recorded optical flashes on the nightside of the planet \cite{cook79,smith79}, while the plasma wave instrument measured low-frequency whistler waves, which on Earth are associated with lightning \cite{gurnett79}. Optical imaging instruments on subsequent missions have also made observations of lightning, including Voyager 2 ISS \cite{borucki92}, Galileo SSI \cite{little99}, Cassini ISS \cite{dyudina04} and New Horizons LORRI \cite{baines07}. In addition to these visible-light observations, the lightning and radio emissions detector (LRD) on the Galileo entry probe also recorded radio emission with a similar waveform to Earth lightning discharge \cite{rinnert98}.

The most recent spacecraft to have detected lightning in Jupiter's atmosphere is the Juno mission, which has been in orbit around Jupiter since 2016 \cite{bolton10}. Three Juno instruments have reported observations of lightning. The Microwave Radiometer (MWR) detected almost 400 lightning events in the 2016--2018 time period \cite{brown18}. These detections were in the form of discrete impulses, measured at a frequency of 600 MHz and were interpreted as lightning sferics, broadband electromagnetic impulses that result from lightning discharge. In addition to these sferics, the radio and plasma wave instrument (Waves) has detected thousands of low-frequency (\textless 20 kHz) whistler waves \cite{kolmavsova18}. On 11 occasions, sferics and whistler waves were observed concurrently by MWR and Waves, marking the first time that lightning in Jupiter's atmosphere has been simultaneously observed by multiple instruments \cite{imai18}. The third instrument that has so far reported lightning observations is the Stellar Reference Unit (SRU), a low light visible imager designed for attitude determination. As with previous optical observations, the SRU lightning detections consist of bright flashes on the planet's nightside \cite{becker20}.

Lightning observed in Jupiter's atmosphere has generally been thought to originate in the planet's water cloud layer, in an analogous manner to intra-cloud lightning on Earth \cite{levin83}. \citeA{borucki86} found that the Voyager 1 optical images were best modelled when the lightning was assumed to occur within a cloud at 5 bar, the estimated location of the water cloud. During the Galileo observations, lightning on the nightside was directly linked to a storm that was imaged on the dayside; analysis of the stormcloud suggested it was located at \textgreater 4 bars \cite{little99}. \citeA{kolmavsova18} also suggested that the lightning observations made by the Juno Waves instrument are likely to originate in Jupiter's water cloud, based on the similar flash density to terrestrial lightning. However, recent observations by the Juno SRU have shown that the lightning processes on Jupiter are more complex than previously thought. \citeA{becker20} found that some of the lightning observed by the SRU occurs at pressures of 1.4--1.9 bar, where pure liquid water cannot exist. Instead, they suggest that a mixed ammonia-water liquid plays an important role in generating this `shallow' lightning. 

While lightning has been observed many times on both Jupiter and other planets in the solar system \cite{zarka86,dyudina10}, there are other atmospheric electricity phenomena that have thus far only been observed in the Earth's atmosphere. Transient Luminous Events (TLEs) are large-scale bright events that occur in the Earth's upper atmosphere and are linked to the electrical activity in an underlying storm system \cite{pasko10}. After many years of eyewitness accounts, a serendipitous observation of a TLE was first recorded in 1989 \cite{franz90}. Since that time, there have been numerous observation campaigns from both the ground \cite<e.g.>{kanmae07,gordillo18} and from space-based instruments \cite<e.g.>{chern03,sato15}. These observations have allowed different types of TLEs to be identified, including elves, sprites, sprite halos and blue jets \cite{pasko10}. Each of these has a different visual appearance and is driven by a different production mechanism. 

In this paper, we report on bright, lightning-like flashes observed by the ultraviolet spectrograph (UVS) on Juno. This is the first time that a Jovian lightning-like phenomenon has been observed in the ultraviolet. These bright flashes are notably different from previous observations of lightning, because Jupiter's atmosphere is optically thick in the ultraviolet, so observations cannot probe beneath the 100-mbar level \cite{vincent00}. Any bright flashes detected by UVS therefore must occur much higher in the atmosphere than both the water cloud and the `shallow' lightning observed by \citeA{becker20}, and must be driven by a different mechanism than intra-cloud discharge. In this paper, we conclude that these bright flashes are consistent with TLEs in Jupiter's upper atmosphere, a phenomenon that has been predicted for Jupiter \cite{yair09} but has not been previously observed. 

In Section~\ref{sec:observations}, we describe the Juno UVS instrument and the method of detection of the bright flashes. Section~\ref{sec:results} describes the analysis of the bright flashes: their spatial extent, lightcurves and spectra. Finally, Section~\ref{sec:discussion} discusses these results and compares them to theoretical predictions for TLEs in Jupiter's atmosphere. 

\section{Observations}
\label{sec:observations}

\subsection{Juno UVS}
\label{sec:juno_uvs}

The Ultraviolet Spectrograph (UVS) is a far-ultraviolet imaging spectrograph on NASA's Juno mission \cite{gladstone17}. UVS covers the 68-210 nm spectral range with a spectral resolution that varies between 1.3 nm and 3.0 nm depending on the position along the instrument's slit \cite{greathouse13}. The primary scientific goal of the UVS instrument is to characterize the morphology, brightness, and spectral characteristics of Jupiter's far-ultraviolet auroral emission and this spectral range includes the H Lyman series and the Lyman, Werner, and Rydberg band systems of H\textsubscript{2}.

Juno is a spin-stabilized spacecraft with a rotation period of $\sim$30 seconds. The UVS instrument slit is nominally parallel to the spacecraft's spin axis, but can be pointed away by up to $\pm$30$^{\circ}$ by using the scan mirror. Photons that enter the instrument slit register as a pulse on the microchannel plate detector. High energy electrons and ions that are present in Jupiter's high radiation environment also register as individual pulses and add noise to the UVS data. UVS data are recorded in a pixel-list time-tagged format; for each photon detection, the x and y position on the detector and the time of the detection are recorded. The x position provides the wavelength of the photon. The y position provides the position along the slit, and this can be combined with the spin phase of the spacecraft at the time of observation in order to assign a position on the planet to each photon. This geometric information can then be used to produce spatial maps of the ultraviolet radiation \cite<e.g.>{gladstone17b}. The UVS slit is made up of two wide segments on either side of one narrow segment, as shown in Figure~\ref{fig:single_spin}. The wide slit has an angular width of 0.2$^{\circ}$, which means that a point source is within the field of view for 17 ms as the spacecraft rotates, and the narrow slit has a width of 0.025$^{\circ}$. The FWHM across the slit is larger than the actual slit size: for the wide part of the slit, it is $\sim$0.25$^{\circ}$ and for the narrow part of the slit it is $\sim$0.2$^{\circ}$ \cite{greathouse13}.

The Juno spacecraft is in a highly elliptical orbit around Jupiter, and UVS obtains high spatial resolution observations of Jupiter for several hours on either side of each perijove (PJ), the point of closest approach to the planet. Data acquisition is paused when the spacecraft passes through regions of Jupiter's radiation belts where the background count rate from high energy electrons and ions overwhelms the count rate from UV photons~\cite{kammer18}. As Juno's orbit precesses over the course of the mission, the PJ location has moved further north. This means that observations of the north polar region have a higher spatial resolution, but are obtained over a shorter time period and have higher background radiation. Observations of the south pole are decreasing in spatial resolution, but are being obtained over a long time period and are more clear of radiation \cite{gladstone19}.

\subsection{Observations of bright flashes}

\begin{figure}
    \centering
    \includegraphics[width=8cm]{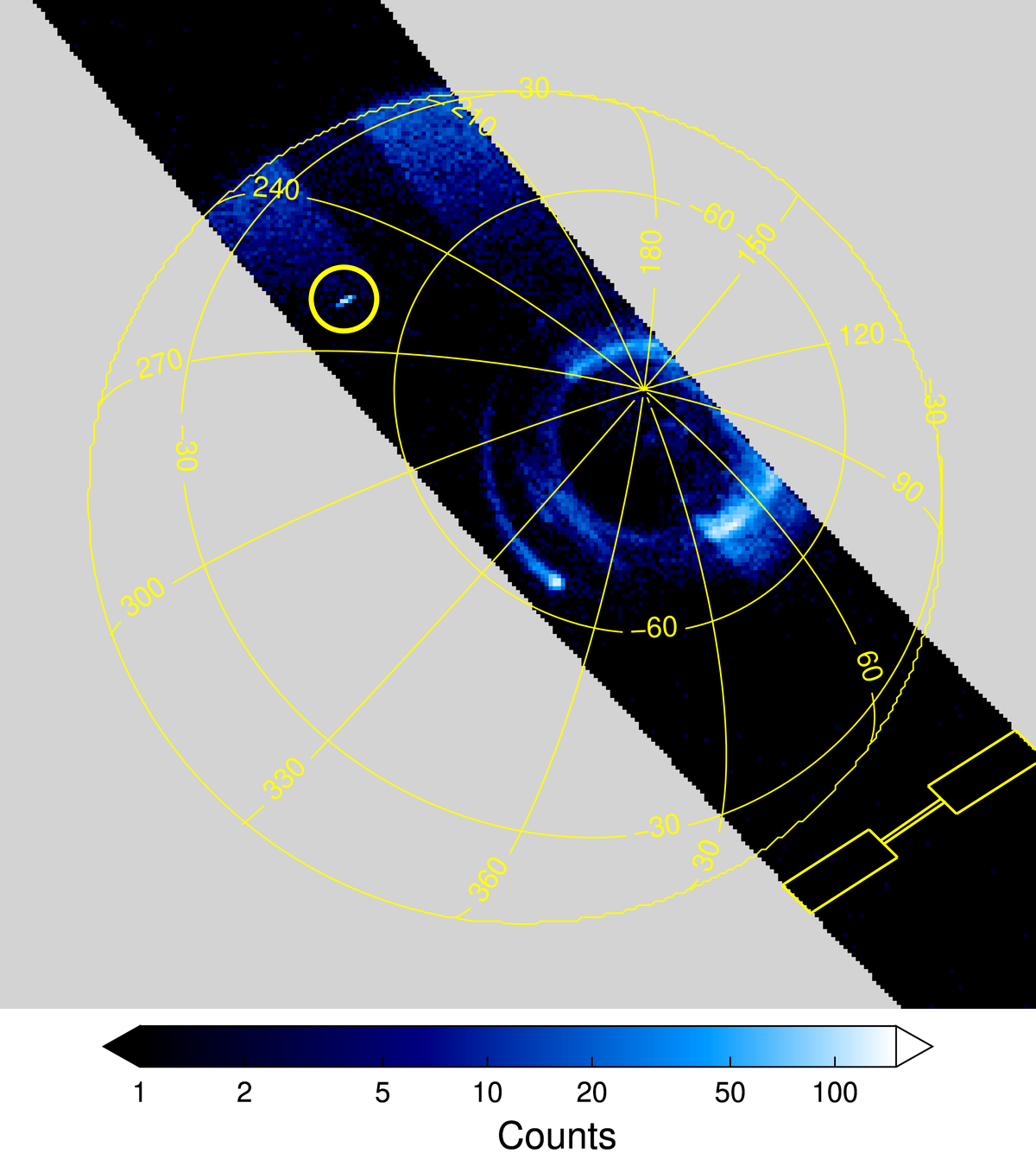}
    \caption{A spatial map of Juno UVS measurements during a single spacecraft spin at PJ26 (10 April 2020). The color scale shows the number of photons counts measured. The shape of the instrument slit is shown in yellow in the lower right corner (widths of the both the wide and narrow slits increased by a factor of 5 for clarity). A bright flash is highlighted by the yellow circle. This flash is also shown in Figure~\ref{fig:images}(j).}
    \label{fig:single_spin}
\end{figure}

The Juno UVS team produces spatial maps of Jupiter for each spin of the spacecraft during each perijove. An example maps for a single spin during PJ26 (10 April 2020) is shown in Figure~\ref{fig:single_spin}. The image swath is built up as the instrument slit sweeps over the planet as the spacecraft rotates. Figure~\ref{fig:single_spin} is centered on Jupiter's south pole, and the data were obtained as the spacecraft was moving away from the planet.

The brightest features in Figure~\ref{fig:single_spin} are the southern auroral oval and the Io footprint, both of which are within 30$^{\circ}$ of the pole. During this spin, another bright feature was detected at a (planetocentric) latitude of 51.0$^{\circ}$S and a longitude of 258.7$^{\circ}$W; this bright feature is highlighted by the solid yellow circle in Figure~\ref{fig:single_spin}. Unlike the auroral emission, this bright spot was very transient. It was not present in the previous or subsequent spins ($\pm$30 s), and as discussed in the following sections, the very short extent in the spin direction indicates that the bright spot is even more short-lived than that.

Data from all perijoves were analyzed to search for comparable short-lived bright flashes. Photon detections within a localized ($\pm$0.4$^{\circ}$) region along the slit were binned into a 5-ms intervals. Bin X was flagged if its count rate was 15 times higher than both bin X-1 and bin X+2, and was also at least 50. The photon detection histrograms for the flagged events were then manually inspected; in some cases, the spike in photon detections was due to auroral emission (evident from the spatial location and a high average count rate over a \textgreater20 ms time interval) or bar code radiation events~\cite<evident from abrupt changes in the count rate that affect the entire slit equally,>{bonfond18}. Through this process, we found a total of 11 events that consist of short-lived, spatially-localized bright flashes. These 11 events are listed in Table~\ref{tab:observations} and are analyzed in Section~\ref{sec:results}.

\begin{landscape}

\begin{table}
\begin{center}
\begin{tabular}{|
>{\raggedright\arraybackslash}p{0.45cm} | 
>{\raggedright\arraybackslash}p{1.65cm} |
>{\raggedright\arraybackslash}p{1.35cm} |
>{\raggedright\arraybackslash}p{1.1cm} |
>{\raggedright\arraybackslash}p{1.1cm} |
>{\raggedright\arraybackslash}p{1.3cm} |
>{\raggedright\arraybackslash}p{1.9cm} |
>{\raggedright\arraybackslash}p{1.9cm} |
>{\raggedright\arraybackslash}p{1.3cm} |
>{\raggedright\arraybackslash}p{1.5cm} |
>{\raggedright\arraybackslash}p{1.6cm} |
>{\raggedright\arraybackslash}p{1.6cm} |
} 
\hline
 & \bf{Date} & \bf{Time} & \bf{Lat ($^{\circ}$)} & \bf{Lon ($^{\circ}$)} & \bf{Solar zenith angle ($^{\circ}$)} & \bf{Wind shear} & \bf{Spacecraft distance (km)} & \bf{Slit} & \bf{Max width (km)} & \bf{Duration (ms)} & \bf{Energy (J)} \\
\hline
(a) & 2016-08-27 & 14:35:15 & -43.2 & 353.7 & 99 & Cyclonic & 172,000 & Wide & 750 & 0.8 & $2.2\times10^{7}$ \\
(b) & 2016-08-27 & 14:57:16 & -43.3 & 350.2 & 88 & Cyclonic & 201,000 & Wide & 1280 & 1.6 & $2.0\times10^{8}$ \\
(c) & 2017-09-02 & 00:34:16 & -51.2 & 236.3 & 103 & Cyclonic & 255,000 & Wide & 1350 & 1.3 & $2.1\times10^{8}$ \\
(d) & 2017-09-02 & 00:39:48 & -43.0 & 263.2 & 119 & Cyclonic & 269,000 & Wide & 1310 & 1.3 & $1.9\times10^{8}$ \\
(e) & 2017-12-16 & 17:09:14 & 34.0 & 165.7 & 137 & Cyclonic & 103,000 & Wide & 660 & 2.5 & $1.8\times10^{7}$ \\
(f) & 2018-09-07 & 05:06:16 & -27.4 & 248.8 & 93 & Cyclonic & 367,000 & Narrow & 1810 & 1.9 & $1.3\times10^{9}$ \\
(g) & 2019-05-29 & 09:17:24 & -61.6 & 222.5 & 110 & Anticyclonic & 127,000 & Wide & 810 & 0.8 & $4.4\times10^{7}$ \\
(h) & 2019-12-26 & 19:03:20 & -66.3 & 166.9 & 73 & Cyclonic & 133,000 & Wide & 920 & 0.1 & $5.3\times10^{7}$ \\
(i) & 2020-04-10 & 13:00:42 & 52.8 & 216.1 & 123 & Cyclonic & 72,000 & Wide & 490 & 1.3 & $2.8\times10^{6}$ \\
(j) & 2020-04-10 & 17:24:35 & -51.0 & 258.7 & 86 & Cyclonic & 338,000 & Wide & 1930 & 1.4 & $5.0\times10^{8}$ \\
(k) & 2020-07-25 & 10:37:13 & -69.0 & 239.9 & 106 & - & 385,000 & Wide & 2180 & 1.4  & $1.9\times10^{8}$ \\
\hline
\end{tabular}
\end{center}
\caption{11 bright flashes observed by Juno UVS. The times are given to the closest second and are as measured on the spacecraft. Latitudes are planetocentric and longitudes are System III West. The wind shear describes whether the winds are cyclonic or anticyclonic at the latitude of the observations (observation (k) occured too close to the pole for the wind shear to be measured). The spacecraft distance is the distance from the bright flash to the spacecraft. The slit column indicates whether the observation was made in the wide part of the slit or the narrow part of the slit. The widths are discussed in Section~\ref{sec:images}, the durations are discussed in Section~\ref{sec:lightcurves} and the energies are discussed in Section~\ref{sec:energy}.}
\label{tab:observations}
\end{table}

\end{landscape}

\section{Results}
\label{sec:results}

\subsection{Images}
\label{sec:images}

\begin{figure}
    \centering
    \includegraphics[width=10cm]{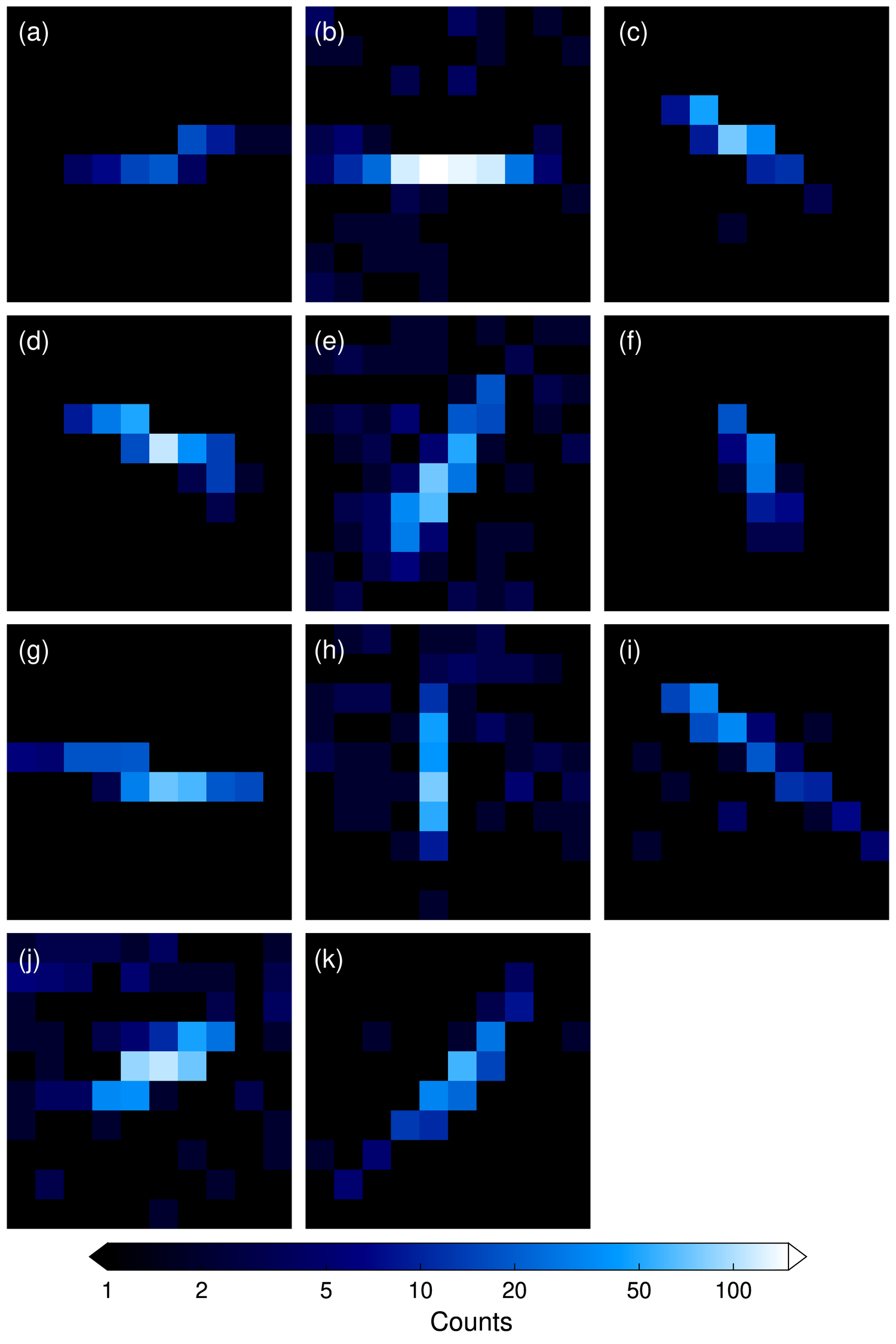}
    \caption{Images of 11 bright flashes observed by Juno UVS. The color scale shows the number of photons counts measured. Each image has dimensions 1$^{\circ}\times$1$^{\circ}$ on the sky. Figure (j) shows the same bright flash as Figure~\ref{fig:single_spin}.}
    \label{fig:images}
\end{figure}

Figure~\ref{fig:images} shows images of the 11 events listed in Table~\ref{tab:observations}. As described in Section~\ref{sec:juno_uvs}, Juno UVS images are generated by the instrument slit sweeping across the planet as the spacecraft rotates. These images were made by mapping the detected photons into 0.1$^{\circ}\times$0.1$^{\circ}$ bins on the sky and each image is 10$\times$10 bins (1$^{\circ}\times$1$^{\circ}$). As they have been mapped onto the sky, the images each have different orientations relative to the slit. In each case, the slit direction is parallel to the streak direction, and the spin direction is approximately perpendicular ($\pm$30$^{\circ}$) to the streak direction. For example, in Figure~\ref{fig:images}(b) the slit is oriented horizontally and in Figure~\ref{fig:images}(h) the slit is oriented vertically. As noted in Table~\ref{tab:observations}, some of the observations ((b), (h), (j)) were made on the dayside of the planet, but the bright flashes are still clearly visible above the background reflected sunlight.

We used a 2-d Gaussian fitting routine \cite{markwardt09} to measure the sizes of the bright regions in Figure~\ref{fig:images}. In the across-slit direction, the FWHM is less than 0.1$^{\circ}$ in all cases. For a constant point source, the measured FWHM across the slit is 0.2--0.3$^{\circ}$ \cite{greathouse13}. The fact that the measured FWHM is narrower suggests a temporal effect, and this is explored in more detail in Section~\ref{sec:lightcurves}. In the along-slit direction, the measured FWHM is 0.33$\pm$0.05$^{\circ}$. When we consider the distance to the spacecraft during each of these measurements (Table~\ref{tab:observations}), this equates to widths of 500--2200 km. However, a FWHM of $\sim$0.3$^{\circ}$ is also consistent with the point spread function of the instrument \cite{greathouse13}, so the source regions could also be considerably smaller than these maximum width values.

These images provide evidence that these observations were due to UV photons from Jupiter, rather than the result of radiation noise. If the hundreds of detected events measured by the instrument were due to a sudden increase in the number of background radiation counts, we would expect the counts to be distributed randomly along the slit, as shown in \citeA{bonfond18}, rather than clustered spatially in a Gaussian shape along the slit.

\subsection{Lightcurves}
\label{sec:lightcurves}

\begin{figure}
    \centering
    \includegraphics[width=15cm]{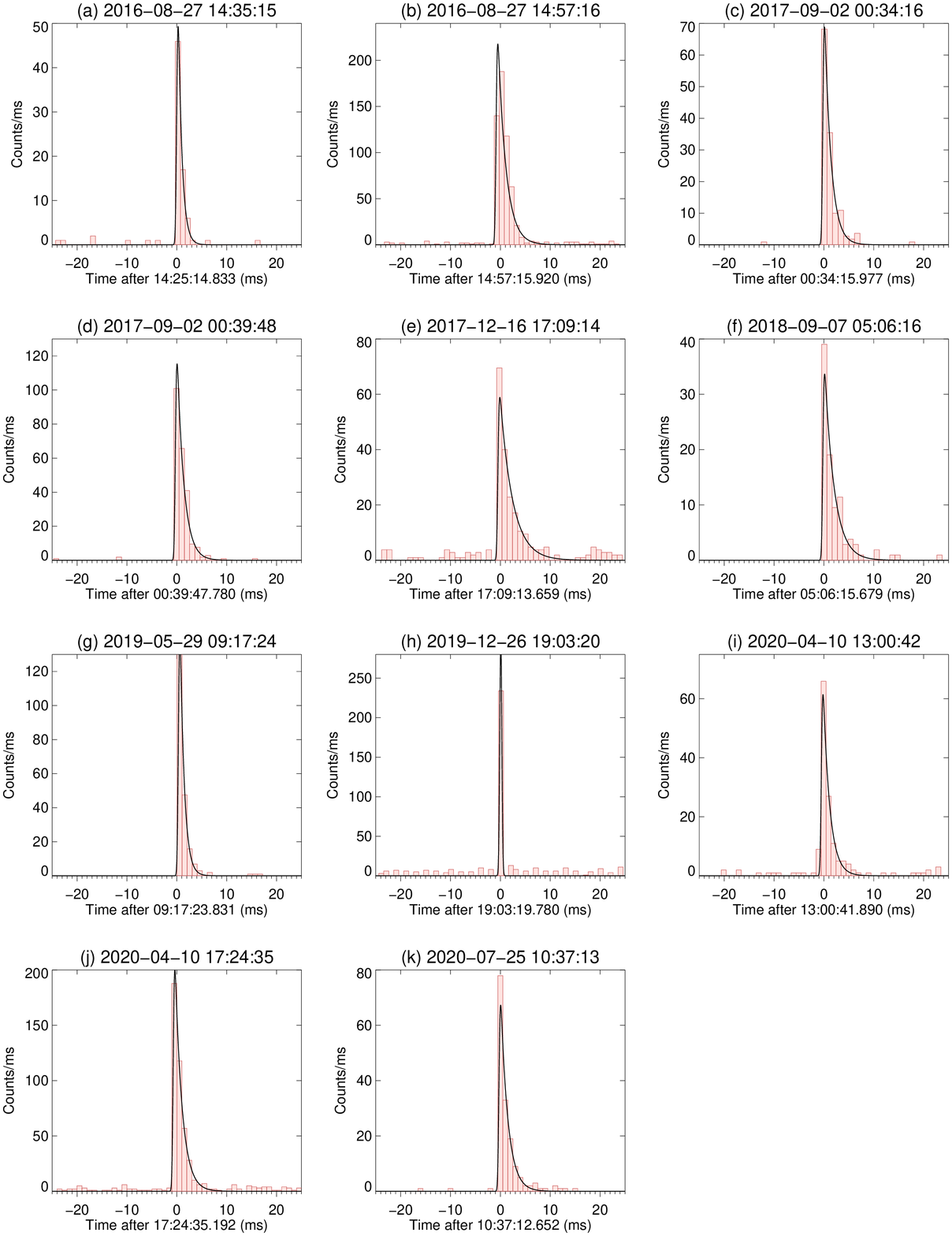}
    \caption{Lightcurves of 11 bright flashes observed by Juno UVS. A histogram of the photon detections is shown in red; bin widths were chosen to produce a smooth distribution and vary slightly between the different frames. A fitted Gaussian-exponential convolution is shown in black for each event.}
    \label{fig:lightcurves}
\end{figure}

Figure~\ref{fig:lightcurves} shows the lightcurves for the 11 events listed in Table~\ref{tab:observations}. Each frame shows the rate of photon detection as a function of time, as the UVS slit scanned over the planet. Because of the finite slit width, a steady point source will remain within the field-of-view of the wide slit for 17 ms (2 ms for the narrow slit). 10 out of the 11 events listed in Table~\ref{tab:observations} were observed in the wide slit, and yet the events shown in Figure~\ref{fig:lightcurves} all have shorter durations than 17 ms. This immediately suggests that the source has a brightness that varies on \textless 17 ms timescales.

The shapes of these events appear to be pulses that decay exponentially with time. We fit the distributions using the convolution of a Gaussian with an exponential decay. The width of the Gaussian was held fixed at 0.2 ms, the value that provided the collective best fit for all 11 events. With this Gaussian width, the fitted exponential decay times are listed in Table~\ref{tab:tles}.

The lightcurves shown in Figure~\ref{fig:lightcurves} may not represent the complete lightcurve for each event; either the beginning or the end of the lightcurve could fall outside the 17 ms interval during which the source is observable. Figure~\ref{fig:lightcurves}(h) has a noticeably narrower shape than the other figures and the fitted decay times is very short (0.1 ms). One possibility may be that the pulse onset occurred immediately before the slit moved off the source, so the decay phase was not captured. For the remaining 10 events, the mean decay time is 1.4$\pm$0.5 ms.

\subsection{Energy}
\label{sec:energy}

We estimated the total energy emitted by the bright source region as follows, based on \citeA{hue19}:

\begin{equation}
    E = f \times \sum_{\lambda=155}^{162} 4\pi d^{2}\frac{hc}{\lambda} \sum_{x,y}C_{x,y}^{\lambda}\Omega_{x,y}
\label{eq:energy}
\end{equation}

As described in Section~\ref{sec:images}, we produce UV maps by binning recorded photons into 0.1$^{\circ}\times$0.1$^{\circ}$ bins on the sky. In Equation~\ref{eq:energy}, these bins are labelled with x and y coordinates. We also bin the photons into 0.6 nm wavelength bins, in order to produce a three-dimensional image cube, where $C_{x,y}^{\lambda}$ is the number of photon counts in a wavelength bin $\lambda$, per solid angle, per detector area.

To convert this image cube into the total energy, we first multiply $C_{x,y}^{\lambda}$ by the solid angle of each x,y bin (0.1$^{\circ}\times$0.1$^{\circ}$) and then sum over the small spatial region containing the bright flash (Figure~\ref{fig:images}). $h$ and $c$ are the Planck constant and the speed of light, and are used to convert the photon detections into energy. $d$ is the distance between the bright flash and the spacecraft when the observations were made, and is used to convert between radiant exposure and radiant energy, assuming photons are emitted isotropically.

In order to calculate the total H\textsubscript{2} emission, we follow the method described in \citeA{bonfond17}. We first sum over photons in the 155--162 nm spectral range, a wavelength range selected because there is negligible atmospheric absorption in the upper atmosphere (see Section~\ref{sec:spectra} for a discussion about the source altitude). Based on a synthetic spectrum, this summed value is then multiplied by $f=8.1$ in order to scale it to the whole H\textsubscript{2} Lyman and Werner bands UV spectrum.

The final H\textsubscript{2} emission energies are in the 10\textsuperscript{6}--10\textsuperscript{9} J range (see Table~\ref{tab:observations}). The systematic and random errors are at least 20\% \cite{gerard19}, and the calculated values are also very dependent of the assumption of isotropic emission.


\subsection{Spectra}
\label{sec:spectra}

\begin{figure}
    \centering
    \includegraphics[width=15cm]{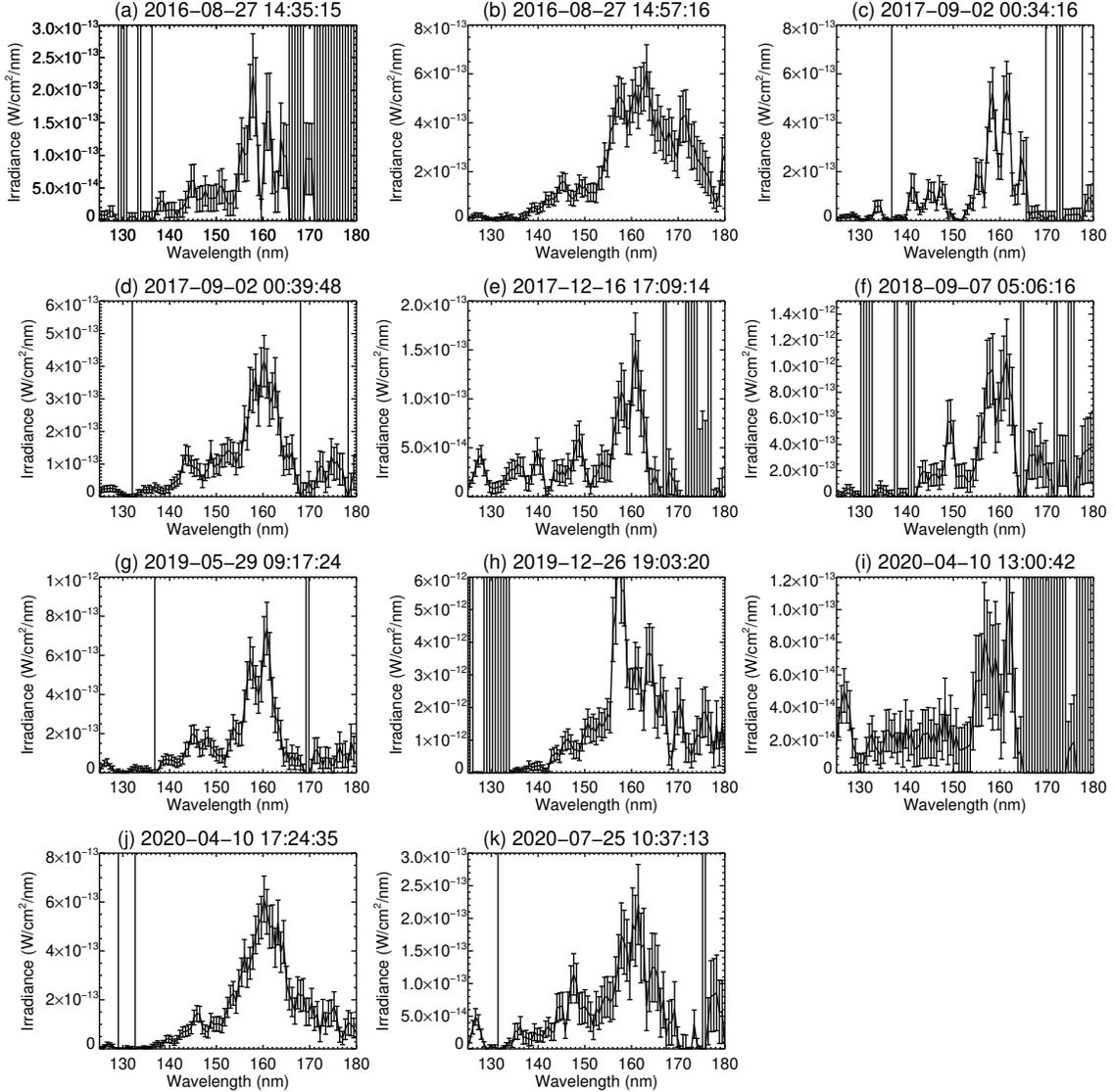}
    \caption{Spectra of 11 bright flashes observed by Juno UVS. Error bars combine shot noise and radiation noise.}
    \label{fig:spectra}
\end{figure}

Figure~\ref{fig:spectra} shows the spectra for the 11 events listed in Table~\ref{tab:observations}. These spectra were calculated in a similar manner to the total energy calculation in Section~\ref{sec:energy}. To convert from radiant energy density to spectral irradiance, we divided each spectrum by the exponential decay times calculated in Section~\ref{sec:lightcurves}. The final spectra are presented in terms of spectral irradiance at the spacecraft location i.e. W/cm\textsuperscript{2}/nm. The spectra have been smoothed using a 2-nm boxcar and they have been background-subtracted to limit the effect of reflected sunlight at the longest wavelengths. The error bars in Figure~\ref{fig:spectra} were calculated combining shot noise (a signal-to-noise ratio of $\sqrt{N}$, where $N$ is the number of photon counts recorded) and radiation noise (calculated from the average count rate at \textless80 nm, where we expect the counts to all be due to radiation).

\begin{figure}
    \centering
    \includegraphics[width=10cm]{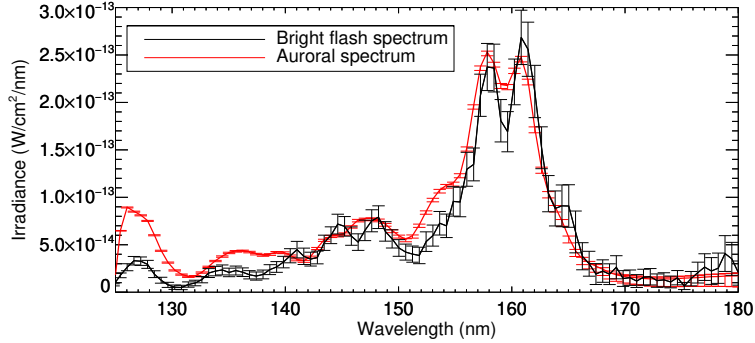}
    \caption{The averaged spectrum (black) of the 7 cleanest Juno UVS bright flash observations ((a), (c), (d), (e), (g), (i) and (k) from Figure~\ref{fig:spectra}). For comparison, a spectrum of Jupiter's aurora as seen by Juno UVS is shown in red (scaled to match the peak irradiance).}
    \label{fig:lightning_spectrum}
\end{figure}

Individually, each of the spectra shown in Figure~\ref{fig:spectra} are somewhat noisy, but it is clear that they do share some broad spectral features, particularly a double peak at 160 nm. In order to conduct further analysis, we combine spectra (a), (c), (d), (e), (g), (i) and (k) into a single spectrum with a higher signal to noise ratio. These are the cleanest spectra, as the observations were made within the wide slit where the signal to noise ratio is higher and they were made on the nightside of the planet, so there is no reflected sunlight. This combined spectrum is shown in black in Figure~\ref{fig:lightning_spectrum}.

Figure~\ref{fig:lightning_spectrum} shows that the bright flash spectrum has a strong double peak at 160 nm, with the irradiance decreasing at both shorter and longer wavelengths. This peak matches the location of a double peak in the H\textsubscript{2} Lyman band that is prominent in observations of Jupiter's auroral emission \cite{gustin04}, immediately suggesting that H\textsubscript{2} emission also plays a significant role in the bright flash spectrum. This spectrum conclusively demonstrates that the bright flashes are not due to radiation noise, as counts from radiation would be randomly distributed in the spectral direction and would produce a featureless spectrum that varies with the effective area of the instrument~\cite{hue19b}.

In order to compare the possible H\textsubscript{2} emission in the bright flash spectrum with the auroral H\textsubscript{2} emission, the red line in Figure~\ref{fig:lightning_spectrum} shows an auroral spectrum from Jupiter, obtained by combining wide-slit Juno UVS data from many perijoves. This auroral spectrum has been binned in the same manner as the bright flash spectrum, although we note that the UVS spectral resolution varies along the slit and therefore the two spectra will not have exactly the same spectral resolution. The auroral spectrum has also been scaled to match the peak irradiance of the bright flash spectrum. The auroral and bright flash spectra both show the same H\textsubscript{2} emission peak at 160 nm, and have a similar spectral shape in the 155--180 nm range. However, at shorter wavelengths, the auroral emission from H\textsubscript{2} is brighter. By 125-130 nm, this difference is approximately an order of magnitude. This suggests that the bright flash has higher absorption from gases such as CH\textsubscript{4} and C\textsubscript{2}H\textsubscript{2} which absorb strongly at these shorter wavelengths, in turn suggesting that the bright flash emission originates from deeper in the atmosphere. 

In order to study this further, we used a model of Jupiter's atmosphere to determine how atmospheric absorption affects the shape of H\textsubscript{2} Lyman band emission. The three stratospheric gases that contribute significantly to absorption in the 125--180 nm range are CH\textsubscript{4}, C\textsubscript{2}H\textsubscript{2} and C\textsubscript{2}H\textsubscript{6}. CH\textsubscript{4} absorption cross-sections were obtained from \citeA{chen04} and \citeA{lee01}, C\textsubscript{2}H\textsubscript{2} was obtained from \citeA{smith91} and \citeA{benilan00}, and C\textsubscript{2}H\textsubscript{6} was obtained from \citeA{lee01}. For a given altitude, the column density of each gas was calculated using the atmospheric composition model described by Model C in \citeA{moses05}. The column densities and absorption cross-sections were used to calculate the atmospheric transmission from the given altitude level. This was then multiplied by an H\textsubscript{2} model from \citeA{gustin04} to produce a top-of-atmosphere spectrum, which was then smoothed to match the Juno UVS observations.

\begin{figure}
    \centering
    \includegraphics[width=10cm]{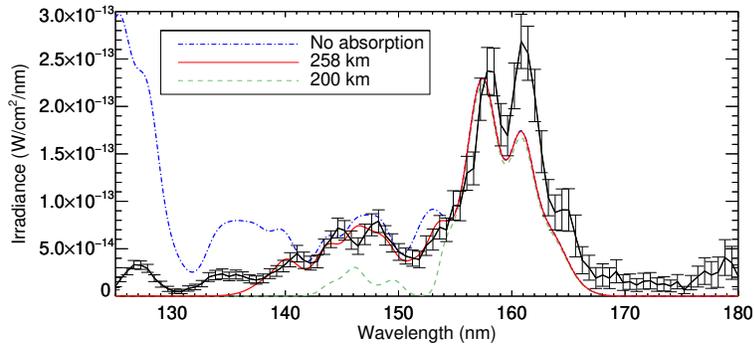}
    \caption{Modeling the bright flash spectrum as H\textsubscript{2} emission attenuated by atmospheric absorption. The best-fit model is shown in red, and corresponds to a source altitude of 258 km above the 1-bar level. For comparison, we show the spectrum of a source with no atmospheric absorption (blue) and a source located deeper in the atmosphere, at 200 km (green).}
    \label{fig:lightning_model}
\end{figure}

This model was used to iteratively fit the observed bright flash spectrum with two free parameters: the source altitude and a scaling factor. The best-fit result is shown by the red line in Figure~\ref{fig:lightning_model} and the corresponding source altitude is 258$\pm$10 km above the 1-bar level (10 \textmu bar). The blue line shows the case where there is no atmospheric absorption, and the green line shows the case where the source is located deeper in the atmosphere, at an altitude of 200 km, so there is more atmospheric absorption. The blue line is far too bright in the 125--140 nm range, while the green line is too dim at 140--150 nm. In contrast, the best-fit red line provides a good fit to the observations throughout the 125--160 nm range. At \textgreater 160 nm, the model does not provide a good fit to the bright flash spectrum; given the similarity between the bright flash spectrum and the auroral spectrum shown in Figure~\ref{fig:lightning_spectrum}, the same discrepancy exists between the model and the auroral spectrum. One possible explanation for this is the presence of H\textsubscript{2} $a$--$b$ continuum emission which peaks in the 200--250 nm range but also extends to shorter wavelengths \cite{pryor98} and is not included in the \citeA{gustin04} model.

\section{Discussion and Conclusions}
\label{sec:discussion}

The events described in Section~\ref{sec:results} and listed in Table~\ref{tab:observations} have the following properties:

\begin{enumerate}[(i)]

    \item They are consistent with point sources (maximum widths of 500--2200 km)
    
    \item Their brightness decays exponentially with time, with a decay time of $\sim$1.4 ms.
    
    \item They are dominated by H\textsubscript{2} emission.

    \item Their total Lyman and Werner band H\textsubscript{2} energy output is in the 10\textsuperscript{6}--10\textsuperscript{9} J range
    
    \item They are located at an altitude of $\sim$260 km above the 1-bar level (at a pressure of 10 \textmu bar).
    
\end{enumerate}

The transient nature and very short decay time of the bright flashes is reminiscent of lightning. However, lightning in Jupiter's atmosphere is generally thought to occur in Jupiter's water cloud layer at $\sim$5 bar \cite{levin83} and even the recently observed `shallow lightning' occurs at pressures \textgreater1.4 bar \cite{becker20}. Ultraviolet observations in contrast cannot probe below 100 mbar \cite{vincent00}, and our analysis shows that the bright flashes are located $\sim$260 above the 1-bar region, which corresponds to $\sim$300 km above the location of the water cloud. It is therefore clear that we are not observing typical cloud-to-cloud lightning strikes. Instead, we suggest that these bright flashes may be upper atmospheric responses to underlying lightning storms, similar to what is seen in the Earth's upper atmosphere.

As discussed in Section~\ref{sec:intro}, Transient Luminous Events (TLEs) are atmospheric electricity phenomena in the Earth's upper atmosphere, which are caused by lightning discharges.  There are several different types of TLEs, including sprites, sprite halos, elves and blue jets. Each of these has different characteristics and a different mechanism, and these are summarized in Table~\ref{tab:tles}. Blue jets emerge directly from the tops of thunderclouds, so are located deeper in the atmosphere than the other types of TLEs, and they also last $\sim$250 ms. For both of these reasons, they seem inconsistent with our observations. However, sprites, sprite halos and elves occur higher in the atmosphere and have shorter durations, so seem to be plausible candidates. 

\begin{landscape}
\begin{table}

\begin{center}
\begin{tabular}{|
>{\raggedright\arraybackslash}p{2cm} | 
>{\raggedright\arraybackslash}p{8cm} |
>{\raggedright\arraybackslash}p{5cm} |
>{\raggedright\arraybackslash}p{2.5cm} |
>{\raggedright\arraybackslash}p{2cm} |
} 
\hline
 \bf{Type} & \bf{Mechanism} & \bf{Shape/Size} & \bf{Altitude} & \bf{Duration} \\ \hline
Blue jets & Discharge between the upper positive charge region in a thundercloud and a negative layer above the cloud & Narrow cone of light extending upwards, $\sim$3 km wide & From cloud tops to 50 km & $\sim$250 ms \\
\hline
Sprites & A cloud-to-ground lightning strike generates a quasi-electrostatic field above the cloud, which accelerates electrons and leads to N\textsubscript{2} emission & Set of luminous columns with a total width of 25--50 km, tendrils extending downwards & 50--90 km & \textgreater5 ms \\
\hline
Sprite halos & As with sprites, they are caused by quasi‐electrostatic thundercloud fields; they sometimes precede the more structured 
sprites & Diffuse disk-shaped emission with a diameter of \textless 100 km & 70--85 km & 2--10 ms \\
\hline
Elves & An intense cloud-to-ground lightning strike causes an upward moving electromagnetic pulse, which heats ambient electrons and leads to N\textsubscript{2} emission & Diffuse and ring-shaped, up to 300 km wide & 75--105 km & \textless1 ms \\
\hline
\end{tabular}
\end{center}
\caption{Types of TLEs observed in the Earth's upper atmosphere. Information obtained from Rodger (1999), Pasko et al. (2010) and Gordillo‐Vázquez et al. (2011).}
\label{tab:tles}
\end{table}

\end{landscape}

Sprites, sprite halos and elves have a similar generation mechanism; in each case, a strong tropospheric lightning strike leads to the acceleration of ambient electrons in the upper atmosphere. In the case of sprites and sprite halos, this is thought to be due to the presence of a quasi-electrostatic field above the cloud tops that accelerates either ambient or high energy electrons produced from cosmic rays, and in the case of elves, this is thought to be due to an upward propagating electromagnetic pulse. In both cases, this leads to the collisional excitation of N\textsubscript{2} molecules, the primary constituent of the Earth's atmosphere \cite{rodger99}.

TLEs have so far only been detected in the Earth's atmosphere, although recent theoretical and laboratory studies have suggested that they may be present on other planets. The possibility was first raised by \citeA{yair09}, who predicted that sprites could occur in Jupiter's upper atmosphere and that their spectra would be dominated by hydrogen emission since that is the primary constituent of Jupiter's atmosphere. This was further confirmed by \citeA{dubrovin10}, who conducted laboratory experiments to show that sprite-like discharges in a Jupiter-like atmosphere are dominated by H and H\textsubscript{2} emission.  \citeA{yair09} developed a quasi-electrostatic model of Jupiter's atmosphere, and using that, they predicted that Jovian sprites would be ignited at an altitude of $\sim$280 km. The possibility of elves in Jupiter's atmosphere has been studied by \citeA{luque14} and \citeA{perez-invernon17}. \citeA{luque14} estimated that for Jupiter the peak photon density for elves would be at an altitude of 250--300 km, and \citeA{perez-invernon17} found that the width of elves in Jupiter's atmosphere could be between $\sim$400 km for horizontal discharge and $\sim$800 km for vertical discharge. For both sprites and elves, the characteristic decay times can vary from sub-millisecond to several milliseconds, depending on the parameters used in the model \cite{luque14, dubrovin14}.

Based on these properties, it is difficult to determine whether our observations are more consistent to be sprites / sprite halos (quasi-electrostatic field generated) or elves (electromagnetic pulse generated). On Earth, the primary distinguishing characteristics are the width, the vertical extent and the duration. Our observed widths are broadly consistent with predictions for all three (although the largest elve predictions slightly exceed our smallest maximum width measurements), and we do not have information on the vertical extents. Our measured duration of $\sim$1.4 ms is similar to the duration of sprite halos of Earth, and is also broadly consistent with the range of values obtained from Jupiter models of sprites and elves.

On Earth, sprites, sprite halos and elves can only be generated in conjunction with a lightning strike. If our observations are indeed one of these TLEs, we would therefore expect there to be a tropospheric lightning flash immediately beforehand. As discussed in Section~\ref{sec:intro}, there are several other instruments on the Juno mission that have previously observed lightning in Jupiter's atmosphere. Unfortunately, none of these instruments observed lightning at the same time as the 11 events we measured with UVS. This is not unexpected, given the manner in which the different observations are all obtained. The Waves instrument detects whistler waves that propagate from the planet to the spacecraft along the magnetic field lines and makes 122-ms observations once per second \cite{kolmavsova18}. The chance of this timing and ``field-of-view'' coinciding with the timing, latitude and longitude of the UVS measurements is low. The MWR instrument has detected many lightning sferics in its 600 MHz channel \cite{brown18}, but the difference in the fields of views of MWR and UVS mean that they cannot observe the same point on the planet at the same time. The field of view of the 600 MHz MWR antenna is 120$^{\circ}$ from the UVS field of view, so as the spacecraft rotates it observes a given point $\sim$20 seconds before and then $\sim$10 seconds after UVS. This time gap is much larger than the expected time gap between lightning flashes and the associated elves or sprites, which is on the order of microseconds and milliseconds respectively \cite{rodger99}. The same issue affects comparisons between UVS and the SRU visible light camera, which has a similar field of view to the 600 MHz MWR antenna.

\begin{figure}
    \centering
    \includegraphics[width=14cm]{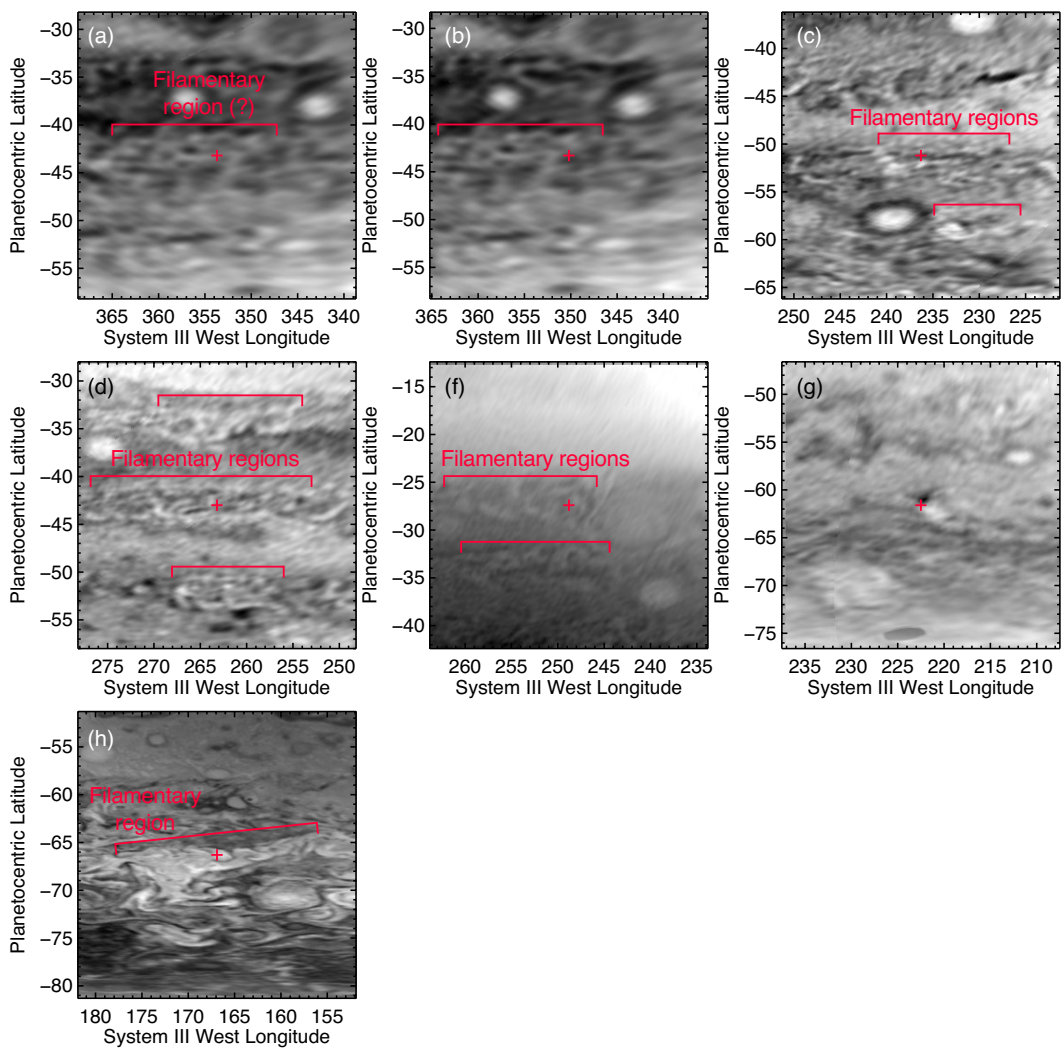}
    \caption{JunoCam images of Jupiter for the locations of 7 of the bright flashes. In each case, the JunoCam data were obtained during the same perijove as the bright flash observation, and they were mostly obtained during the departure sequence as the spacecraft was moving away from the planet. The images shown here were made by combining data from the red and green JunoCam filters, obtained from the PDS (see the Data Availsbility Statement). The locations of the bright flashes are shown by the red crosses and the folded filamentary regions are marked by the lines. The geometry of the orbit means that there is no daylight JunoCam image for the location of flash (e). Processed JunoCam images for (i), (j) and (k) are not yet available. A higher resolution image of the region around flash (f) is shown in Figure~\ref{fig:hst_map}.}
    \label{fig:junocam_map}
\end{figure}

\begin{figure}
    \centering
    \includegraphics[width=9cm]{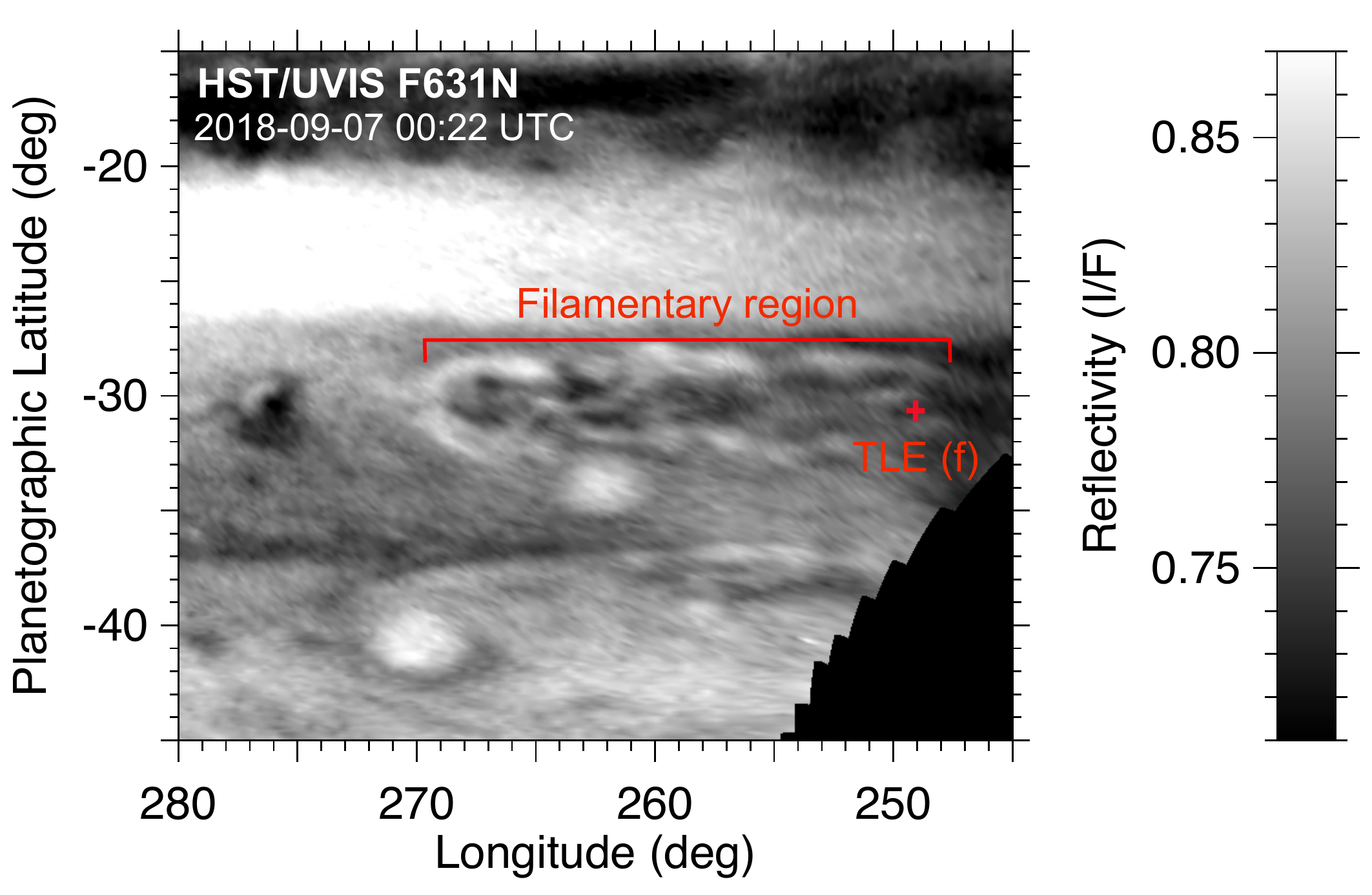}
    \caption{HST WFC3/UVIS image of Jupiter, obtained 5 hours before bright flash (f) was observed. The scale gives the equivalent disk-center reflectivity after correction for limb darkening. The latitude and longitude of bright flash (f) is marked by the red cross.}
    \label{fig:hst_map}
\end{figure}

While there are no direct observations to link the bright flashes with tropospheric lightning, we do note that they occur in regions of the planet where we would expect lightning to be present. Table~\ref{tab:observations} lists whether the wind shear at the latitude of each observation is cyclonic or anticyclonic; vorticities were obtained from Hubble Space Telescope observations \cite{tollefson17,wong20}. Out of the 10 observations for which wind shear could be measured, 9 occurred at latitudes with cyclonic zonal wind shear. This is consistent with the bright flashes being caused by lightning, because lightning on Jupiter is more frequently observed in the cyclonic belts than in the anticyclonic zones \cite{little99,brown18}.

Further evidence comes from Figures \ref{fig:junocam_map} and \ref{fig:hst_map}, which show images of Jupiter obtained using JunoCam and the Hubble Space Telescope respectively. JunoCam is the outreach camera on the Juno mission and it visible-light wide-angle imaging during each perijove \cite{hansen17}. Figure \ref{fig:junocam_map} shows JunoCam images for 7 out of the 11 bright flashes; because of the geometry of the orbit, no dayside images of region near flash (e) were obtained, and processed JunoCam data for flashes (i), (j) and (k) are not yet available from NASA's Planetary Data System (PDS). Figure \ref{fig:hst_map} shows an image obtained using the HST Wide Field Camera 3 within 5 hours of flash (f) and covering the longitude and latitude of the flash \cite{wong20}. Figures \ref{fig:junocam_map} and \ref{fig:hst_map} show that several of the bright flashes occurred within folded filamentary regions, oblong cyclonic regions containing fine-scale filamentary structure \cite{morales02}. Many previous lightning detections have been traced to cyclonic regions \cite{vasavada05,wong20}, or to regions that share the three cloud structure elements of active moist convection: high/thick convective towers, deep water clouds, and cloud-free clearings \cite{gierasch00,imai20}.

As the Juno mission continues, the UVS instrument will continue to be used to search for TLEs in Jupiter's atmosphere. More observations will increase the probability that there is an overlap with the Waves instrument observations, and will also allow us to conduct a statistical analysis of the occurrence rate and spatial distribution of our observations. Any observations made closer in to the planet will also help us to constrain the width of the bright flashes, which in turn will help to discriminate between sprites, sprite halos and elves. 

\section*{Acknowledgements}

We are grateful to NASA and contributing institutions, which have made the Juno mission possible. We thank Thomas Momary for advice on processing the JunoCam images. This work was funded by NASA's New Frontiers Program for Juno via contract with the Southwest Research Institute.

\section*{Data Availability Statement}

The Juno UVS and JunoCam data used in this paper are archived in NASA's Planetary Data System (PDS). The Juno UVS data are available at the PDS Atmospheres Node: https://pds-atmospheres.nmsu.edu/PDS/data/jnouvs\_3001 \cite{trantham14}. The JunoCam data are available at the PDS Imaging Node: https://pds-imaging.jpl.nasa.gov/data/juno \cite{caplinger14}. The HST observations are archived at the Mikulski Archive for Space Telescopes: https://archive.stsci.edu/hlsp/wfcj \cite{wong20}. The data used to produce the figures in this paper are available in \citeA{giles20b}.

\nocite{gordillo11} 


\end{document}